\documentclass[lengthcheck,onecolumn,pre,superscriptaddress,showpacs,floatfix,amsmath,amssymb,aps]{revtex4-2}

\usepackage{graphicx}
\usepackage{dcolumn}
\usepackage{bm}
\usepackage{color}

\usepackage{tikz}
\newcommand{\circled}[1]{\tikz[baseline=(char.base)]{\node[shape=circle,draw,inner sep=1.3pt] (char) {#1};}}

\begin{document}

\preprint{}

\title{Multi-agent rendezvous in fluid flows via reinforcement learning}

\author{Bocheng Li$^{1}$}
\author{Jingran Qiu$^{2,*}$}
\author{Lihao Zhao$^{1,*}$}
\affiliation{$^{1}$AML, Department of Engineering Mechanics, Tsinghua University, Beijing, 100084 
 China.\\$^{2}$Department of Physics, Gothenburg University, Gothenburg, 41296 Sweden.}

\date{\today}

\begin{abstract}
Rendezvous is a critical task for multi-agent systems, requiring agents to coordinate to meet at an unspecified location. However, achieving this in fluid environments presents a challenge, as it remains unclear how agents can exploit underlying fluid kinematics to facilitate convergence. In this study, we adopt a multi-agent reinforcement learning (MARL) approach to develop physics-informed rendezvous strategies in vortical flows. Compared to a naive strategy, where agents navigate toward their counterparts, MARL strategies significantly improve the rendezvous rate. MARL strategies also show transferability across varying vortex intensities, vortex scales, and swarm sizes. By breaking the symmetry of the state-action map, MARL strategy leverages a non-intuitive mechanism that prevents agents from becoming trapped in separate vortices, thereby enhancing rendezvous success. Additionally, a heuristic strategy is extracted from the learned strategy and also outperforms the naive strategy. Furthermore, a theoretical analysis demonstrates that fluid deformation impedes the rendezvous process. Large finite-time Lyapunov exponents identify where fluid effects separate adjacent agents, suggesting that targets should be planned in weak-deformation regions. Our findings reveal the important role that agent-fluid interactions play in multi-agent tasks and highlight the MARL capability to explore swarm intelligence in complex flow environments.
\end{abstract}
\maketitle

\section{Introduction}\label{sec1}

The rendezvous problem is one of the fundamental problems in multi-agent systems, in which agents in a swarm follow a strategy to eventually rendezvous at an unspecified location \cite{linMultiagentRendezvousProblem2003}. In nature, rendezvous is essential for creature mating and information sharing \cite{amirkhaniConsensusMultiagentSystems2022}. For multi-vehicle systems, such as robots and even spacecrafts, rendezvous is essential for payload delivery and logistics \cite{wangModelbasedReinforcementLearning2020,bregerSafeTrajectoriesAutonomous2008}. Meanwhile, the rendezvous is a premise task for realizing swarm intelligence \cite{chasePhysicsSensingDecisionMaking}, such as flocking to efficiently migrate \cite{ganduriSwarmIntelligenceAction2024}, cloak themselves \cite{mirzakhanlooActiveCloakingStokes2020}, etc. Therefore, investigating the rendezvous problem is of significant value for agents across length scales and has drawn much attention in recent years.

Many rendezvous processes take place in fluid environments. The motions of aquatic swimmers \citep{pedleyHydrodynamicPhenomenaSuspensions}, oceanic ships \citep{songMultivehicleCooperationNearly2017}, and balloons \citep{bellemareAutonomousNavigationStratospheric2020}, are significantly influenced by the ambient fluid motions. Fluid dynamics is typically non-linear, which makes traditional policies suboptimal. Few studies have considered flow effects. Sang et al. \cite{sangPathPlanningFormation2024} use a particle swarm optimization algorithm to optimize the rendezvous paths of wave gliders, where only large-scale currents are considered, while small-scale structures create a challenge for the rendezvous task. Zaidi et al. \cite{zaidiAdaptiveActiveDisturbance2022} achieved time-varying rendezvous for the gust-disturbed drones with a leader-follower method, where the wind gust is treated as disturbance since it is much weaker than the propelling ability of drones. Overall, how ambient fluid motions affect the rendezvous task remains unknown.

Three categories of methods have been proposed to study the rendezvous problem, i.e., virtual forces, probabilistic approaches, and artificial evolution methods \citep{bayindirReviewSwarmRobotics2016}. Multi-agent reinforcement learning (MARL) is an evolution method, where neural networks link sensory inputs to actuator outputs, and these networks
evolve with experience \citep{caneseMultiAgentReinforcementLearning2021}. In earlier studies, reinforcement learning (RL) performs well in controlling agents in various flow fields. For instance, ones study the navigation of microswimmers in vortical and turbulent flows to reveal the response mechanism to fluid signals for different goals \citep{colabreseFlowNavigationSmart2017, qiuNavigationMicroswimmersSteady2022,gunnarsonLearningEfficientNavigation2021,alageshanMachineLearningStrategies2020,xuLongdistanceMigrationMinimal2023,liEscapePredatorinducedFlow2025,yangMachineLearningMicro2024,jiaoSensingFlowGradients2025}. Other agents, including the glider \citep{reddyGliderSoaringReinforcement2018}, airship \citep{zhengPathPlanningStratospheric2024}, and stratospheric balloon \citep{bellemareAutonomousNavigationStratospheric2020}, are also effectively controlled to take advantage of wind effects. These works show that RL is capable to utilize physical mechanisms to tackle the navigation problems in flow fields, which inspires us to investigate how a MARL approach can discover rendezvous strategies. To the best of the authors' knowledge, the only study applying MARL to agents in flow fields is that Borra et al. \cite{borraReinforcementLearningPursuit2022} investigated the pursuit-evasion problem of two microswimmers by a decentralized MARL method. Hence, attempts to exploit MARL on different multi-agent problems are meaningful, especially on the current rendezvous problem.

How does ambient fluid motion affect the rendezvous task? Can we find an effective rendezvous strategy by utilizing flow signals in MARL implementation? If so, what is the mechanism? These questions motivate us to devise a training approach based on MARL and carry out numerical experiments to explore and interpret the rendezvous strategies. This work reveals important flow effects on the rendezvous task and provides effective and robust strategies. Our approach to applying MARL to problems with agent-fluid interactions shows promise and provides guidance for further exploitation of MARL to specific agents in complex flow environments.

\section{Methods}
\begin{figure}[h]
      \centering
      \includegraphics[width=0.6\textwidth]{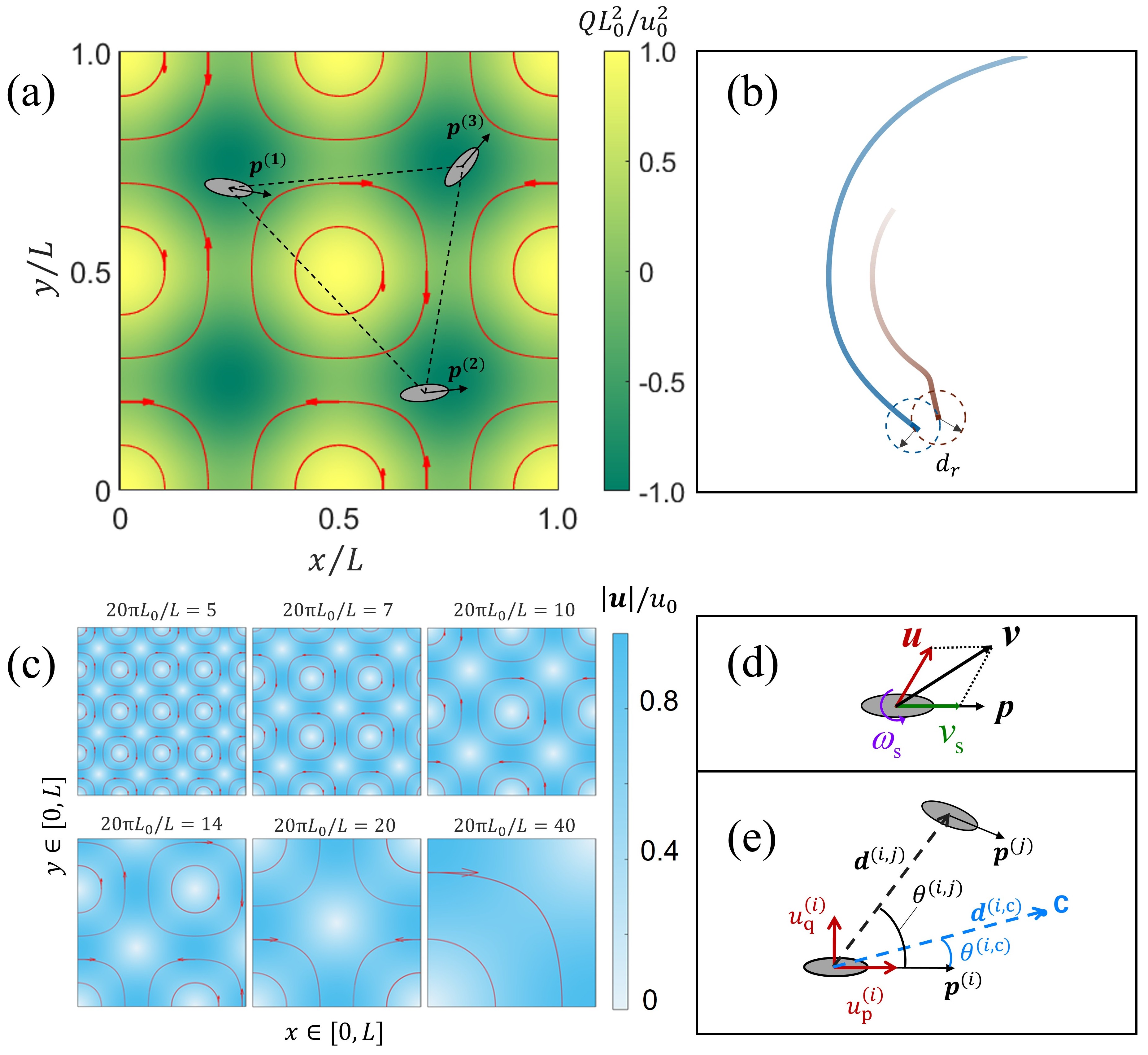}
      \caption{(a) Three agents moving in Taylor-Green Vortices of $L/L_0 = 2\pi$. The black arrow represents the agent orientation $\bm{p}$. The red lines are the streamlines, and the background color indicates the vortex criterion $Q$ value. (b) A rendezvous trajectory of two agents. The radius of the ending circle represents the distance recognized as a successful rendezvous. (c) The Taylor-Green Vortices of different vortex scale $L_0$. The shown region is $\Phi$. The red lines are streamlines, and the background color stands for the flow speed.  (d) The symbols of agent motion, where $v_\text{s},\omega_\text{s}$ represent the propelling and steering speed, respectively. $\bm{v},\bm{u}$ are the velocity of the agent and local fluid, respectively. (e) The state symbols for a swarm. $u_{\text{p}}^{(i)},u_{\text{q}}^{(i)}$ denote the fluid velocity components on the local coordinates of agent $i$. Point 'c' represents the spatial center of this swarm. $\bm{d}^{(i,j)}$ is the vector from position $i$ to position $j$, and $\theta^{(i,j)}$ is the angle between $\bm{d}^{(i,j)}$ and $\bm{p}^{(i)}$.}
      \label{sys}
\end{figure}

In this study, agents are modeled as point-masses, allowing inter-agent collisions to be neglected, and each agent possesses global observability of all other agents \citep{HuettenrauchAdrianNeumann2019_1000118251}. An example swarm consisting of $n = 3$ agents is shown in Fig.~\ref{sys}(a). The swarm is considered to successfully rendezvous when the distance between any two agents is less than a set value $d_r$, as shown in Fig.~\ref{sys}(b). Since the rendezvous process of each swarm is dependent on its initial condition, a statistical indicator is defined to evaluate the effectiveness of a rendezvous strategy as the rendezvous rate $R_\text{r} = N_\text{r}/N$, where $N$ is the number of swarms randomly initialized in region $\Phi$ $(x\in[0,L], y\in[0,L])$, and $N_\text{r}$ is the number of swarms that successfully rendezvous within a time horizon $T_\text{t}$. $R_\text{r}$ is obtained through Monte Carlo method based on numerical simulations. In the following sections \ref{2.1} to \ref{2.4}, fluid-agent interaction models and training details are presented.

\subsection{Flow field}\label{2.1}

We consider a typical vortical flow field, i.e., steady Taylor-Green Vortices (TGV) flow, which contain periodic counter-rotating vortices with intensity $u_0$ and scale $L_0$ \citep{taylorLXXVDecayVortices1923}, as shown in Fig.~\ref{sys}(a). The flow velocity components are 
\begin{equation}
  \begin{gathered}
      u_x = u_0\cos(\frac{x}{L_0})\sin(\frac{y}{L_0}),\\
      u_y = -u_0\sin(\frac{x}{L_0})\cos(\frac{y}{L_0}).
  \end{gathered}
  \label{eq:tgv}
\end{equation}
The maximal flow speed is $u_0=\max{\sqrt{u_x^2+u_y^2}}$. The larger $L_0$ corresponds to larger velocity variation and denser vortices, as shown in Fig.~\ref{sys}(c). Since TGV flow exhibits typical rotation and deformation, we characterize its structural features using the $Q$-criterion, defined as the rotational kinetic energy minus deformation kinetic energy:
\begin{equation}
  \begin{gathered}
      Q =\frac{1}{2}(\Omega_{ij}\Omega_{ij}- E_{ij}E_{ij}) =(\frac{u_0}{L_0})^2 \cos(\frac{x+y}{L_0})\cos(\frac{x-y}{L_0}),
  \end{gathered}
  \label{eq:tgvQ}
\end{equation}
where $\Omega_{ij}, E_{ij}$ represent the components of the rotation matrix and deformation matrix of fluid motion, respectively. The $Q$-criterion is an important diagnostic associated with the rendezvous behavior, which is discussed in the following analysis in Section \ref{sec:dist}.

\subsection{Agent dynamics}\label{2.2}

In the flow field, agents can propel themselves and are also subjected to fluid velocity $\bm{u}$, as shown in Fig.~\ref{sys}(d). An ideal model governs the agent motion \citep{biferaleZermeloProblemOptimal2019}:
\begin{equation}
  \begin{gathered}
      \frac{{\mathrm{d}\bm{x}}}{{\mathrm{d}t}} = \bm{v} = \bm{u} + {v_\text{s}}\bm{p},\\
      \frac{{\mathrm{d}\bm{p}}}{{\mathrm{d}t}} = \bm{\omega}  \times \bm{p}= \omega_\text{s} \bm{e_\text{z}}\times \bm{p}, 
  \end{gathered}
  \label{eq:veh}
\end{equation}
where $\bm{x}$ denotes the agent position, and $\bm{e_\text{z}}$ is the unit vector of the z-axis (normal vector of the plane). The agent propels at a fixed speed $v_\text{s}$ directing to orientation $\bm{p}$, and its steering speed $\omega_\text{s}$ is the action controlled by a strategy, whose magnitude is restricted not to exceed $\omega_\text{m}$. The time scale in this system is defined as $\tau = \omega_\text{m}^{-1}$. The agent trajectories are obtained by integrating \eqref{eq:veh} with Adams-Bashforth scheme.

\subsection{Multi-agent reinforcement learning}\label{2.3}

\begin{figure*}
      \centering
      \includegraphics[width=0.9\textwidth]{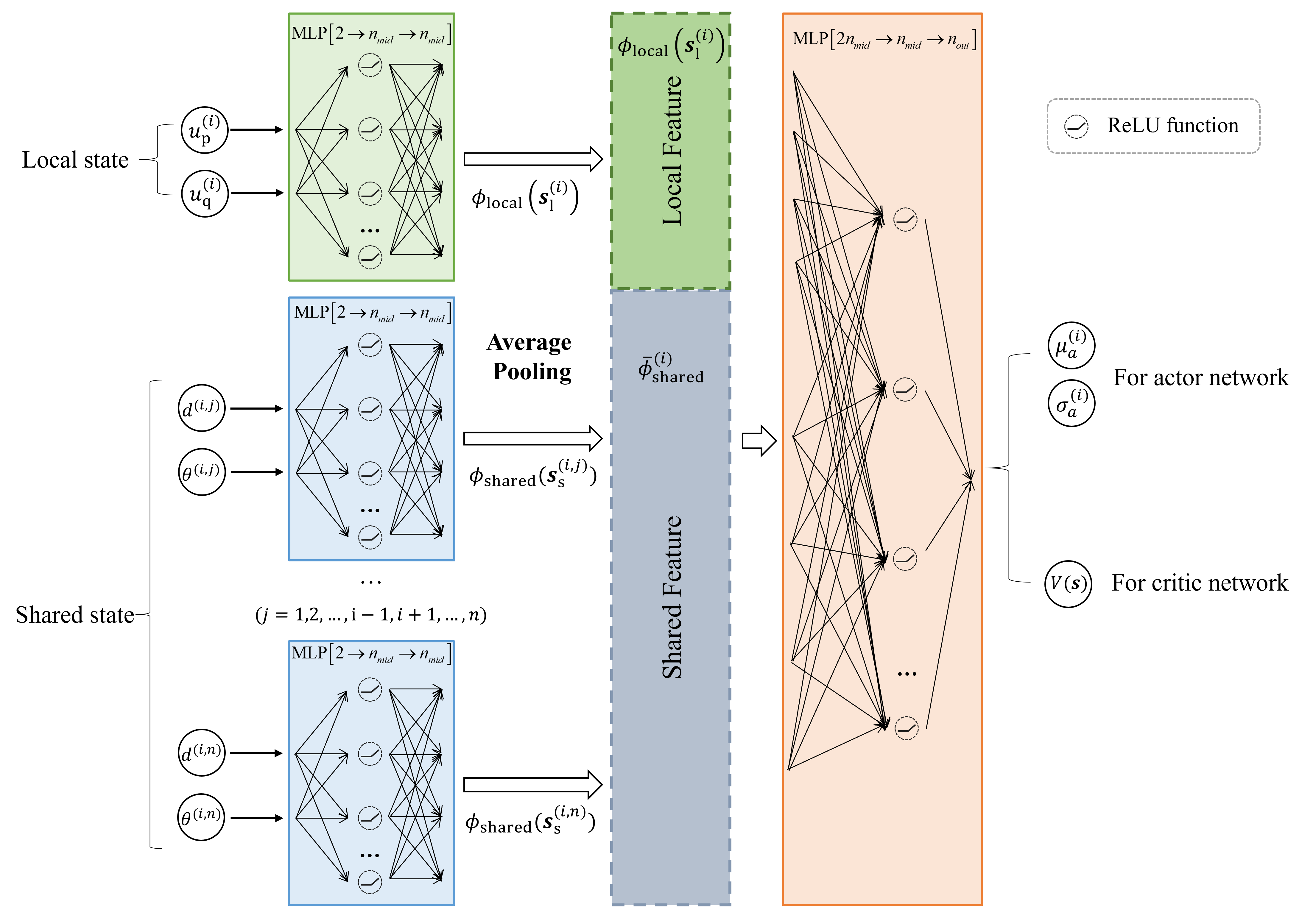}
      \caption{The diagram of DS-PPO network architecture, with the state of the $i$-th agent in the swarm as input. For the actor network, the output variable is the mean and standard deviation of the action distribution $(\mu_\text{a}, \sigma_\text{a})$. For the critic network, the output is the value estimate $V(\bm{s})$ of input states. Each MLP block denotes a two-layer fully connected network with ReLU activation in between, where the numbers of neurons are indicated in the brackets.}
      \label{net}
\end{figure*}

To achieve rendezvous, agents take the action $\omega_\text{s}$ to adjust the propelling direction based on perceived signals following a specific strategy, which can be obtained by MARL training. As illustrated in Fig.~\ref{sys}(e), the signals consist of the local states (flow velocity components $u_{\text{p}}^{(i)}$, $u_{\text{q}}^{(i)}$), and the shared states (relative positions of other agents $d^{(i,j)}$, $\theta^{(i,j)}(j=1,2,...,i-1,i+1,...,n)$, where $d^{(i,j)}=|\bm{d}^{(i,j)}|$). A naive strategy is defined as one where each agent always steers to orient to the center of this swarm, regardless of the flow effects \citep{ribeiroRendezvousAlgorithmMultiagent2020}. For the $i$-th agent, the naive strategy is
\begin{equation}
  \begin{gathered}
\omega_\text{s}^{(i)}= \text{sgn}(\theta^{(i,\text{c})})\omega_\text{m},\\
\theta^{(i,\text{c})} = \cos^{-1} \frac{<\bm{d}^{(i,\text{c})},\bm{p}^{(i)}>}{|\bm{d}^{(i,\text{c})}||\bm{p}^{(i)}|},\\
\bm{d}^{(i,\text{c})} = \frac{\sum_{j=1}^{n}\bm{d}^{(i,j)}}{n}.
  \end{gathered}
\end{equation}

We apply Deep Set Proximal Policy Optimization (DS-PPO) algorithm to explore the rendezvous problem. The network structure is motivated by Deep Sets~\citep{zaheer2017deep,li2021permutation}, which can preserve the permutation invariance of the individuals in a swarm. DS-PPO avoids dimensional explosion as a fully decentralized MARL algorithm, and it adapts to different swarm sizes $n$ because the shared information is processed with a pooling layer. The network architecture is shown in Fig.~\ref{net}.

All agents in the swarm share the same network and parameters.
For the $i$-th agent, the input of the network consists of two parts, the local states $\bm{s}_\text{l}^{(i)}=\{u_{\text{p}}^{(i)},u_{\text{q}}^{(i)}\}$ measured by an agent, and the shared states $\bm{s}_\text{s}^{(i,j)} = \{d^{(i,j)}, \theta^{(i,j)}\} (j=1,2,...,i-1,i+1,...,n)$ that it receives from all individuals in the swarm. The local state is fed to a fully connected network $\phi_\text{local}(\bm{s}_\text{l}^{(i)})$. The shared state is fed to another fully connected network $\phi_\text{shared}(\bm{s}_\text{s}^{(i,j)})$, followed by an average pooling $\overline{\phi}_\text{shared}^{(i)} = \tfrac{1}{n-1}\sum_{j=0,j\neq i}^n{\phi_\text{shared}(\bm{s}_\text{s}^{(i,j)})}$. We note that all the shared states, $\bm{s}_\text{s}^{(i,j)}, j=1, ...,n$, are fed to the same network $\phi_\text{shared}$, which keeps the permutation invariance of the agents in the swarm. Finally, $\phi_\text{local}(\bm{s}_\text{l}^{(i)})$ and $\overline{\phi}_\text{shared}^{(i)}$ are concatenated and fed into the output layer $(\mu_\text{a}^{(i)}, \sigma_\text{a}^{(i)}) = \rho(\phi_\text{local}(\bm{s}_\text{l}^{(i)}),\overline{\phi}_\text{shared}^{(i)})$, where $\mu_a^{(i)}$ and $\sigma_a^{(i)}$ are the mean and standard deviation of the action distribution for this agent. We use the same network architecture for both the actor and critic networks. The hyperparameters of the network architecture are given in Fig.~\ref{net}, where $n_\text{mid}=200$.

The navigational strategy, modeled by the network, can be optimized by training. The training process is to search for a network that maximizes a reward, whose growth represents the rise of the rendezvous rate. The reward is designed to evaluate the action taken at each state. Having the total distance between agent $i$ and other agents $D^{(i)}=\sum_{\substack{j=1,j\neq i}}^{n}d^{(i,j)}$, the reward given to agent $i$ at the $m$-th time step $R^{(i,m)} = D^{(i,m)}-D^{(i,m+1)}$, which means that if the total distance decreases, a positive reward is given in training. 

 The strategy is trained by updating the networks following the standard way of Proximal Policy Optimization~\citep{schulmanProximalPolicyOptimization2017}, which requires the trajectories of agents with data including states, actions and rewards. Here, all the trajectories of individuals in a swarm are used in the training. The training is performed in an episodic way. In each episode, ten swarms are initialized in $\Phi(x,y\in [0,L])$, and the agents move until all swarms successfully rendezvous, or the maximum duration of an episode $T_\text{t}=100\tau$ is reached. The value of $T_\text{t}$ is chosen such that any two-agent swarm using naive strategy has sufficient time to rendezvous in a quiescent flow field. Every $0.025\tau$, the agents decide their actions according to the actor network, and every $5\tau$, the networks are updated once. During updating, $\gamma=0.99$ is used in the calculation of generalized advantage estimation. For better convergence, the learning rate $\alpha$ decays with episode $E$ as $\alpha  = \max\left[0,\alpha _0\left(1-\frac{E}{E _0}\right)\right]$, where hyperparameters are set as $\alpha_0=10^{-4}, E_0=985$. In each case, we train the networks for no more than 1000 episodes. Once training converges (i.e. the total reward in one episode does not change with further training), we evaluate the strategy by setting $\sigma_\text{a} = 0$, so that the action is $\omega=\mu_\text{a}\omega_\text{m}$.

\subsection{Cases and parameters}\label{2.4}
The steering and propelling abilities, i.e., $\omega_\text{m}$ and $v_\text{s}$, are taken as characteristic variables. Parameters that may influence the rendezvous rate $R_\text{r}$ are non-dimensionalized and listed in Table \ref{para}. To understand the effects of flow intensity and scale on rendezvous task, we consider different TGV intensities and scales, $\beta_{u} \in \{0,1,2,3,4,5\}$, $\beta_{l} \in \{5,7,10,14,20,40\}$. Furthermore, we also consider different swarm sizes $n \in \{2,3,4\}$. In order to test the transferability of MARL strategy, we trained the strategy for a fixed $\beta_{l}=10$, and tested it for varying $\beta_{l}$. Hereinafter, we denote each learned MARL strategy as 'N$n$U$\beta_{u}$', representing the optimal model selected from five independent training trials.

\begin{table}[h]
\caption{\label{para}Dimensionless parameters in the rendezvous task.}
\begin{ruledtabular}
  \begin{tabular}{lcc}
    Parameter & Dimensionless & Value\\
    \colrule
    Initial domain length $L$ & $\beta_L = L\omega_\text{m}/v_\text{s}$ & $20\pi$\\
    Rendezvous distance $d_\text{r}$ & $\beta_d = d_\text{r}\omega_\text{m}/v_\text{s}$ & $1$\\
    Motion duration $T_\text{t}$ & $\beta_T = T_\text{t}\omega_\text{m}$ & $100$\\
    TGV intensity $u_0$ & $\beta_u = u_0/v_\text{s}$ & \{0,1,2,3,4,5\}\\
    TGV scale $L_0$ & $\beta_{l} = L_0\omega_\text{m}/v_\text{s}$ & \{5,7,10,14,20,40\}\\
    Swarm size $n$ & $n$ &  \{2,3,4\}\\
    \end{tabular}
    \end{ruledtabular}
\end{table}

\section{Results}
\subsection{Performance of two-agent strategies}\label{sec:2agent}
\begin{figure*}
  \centering
  \includegraphics[width=0.88\textwidth]{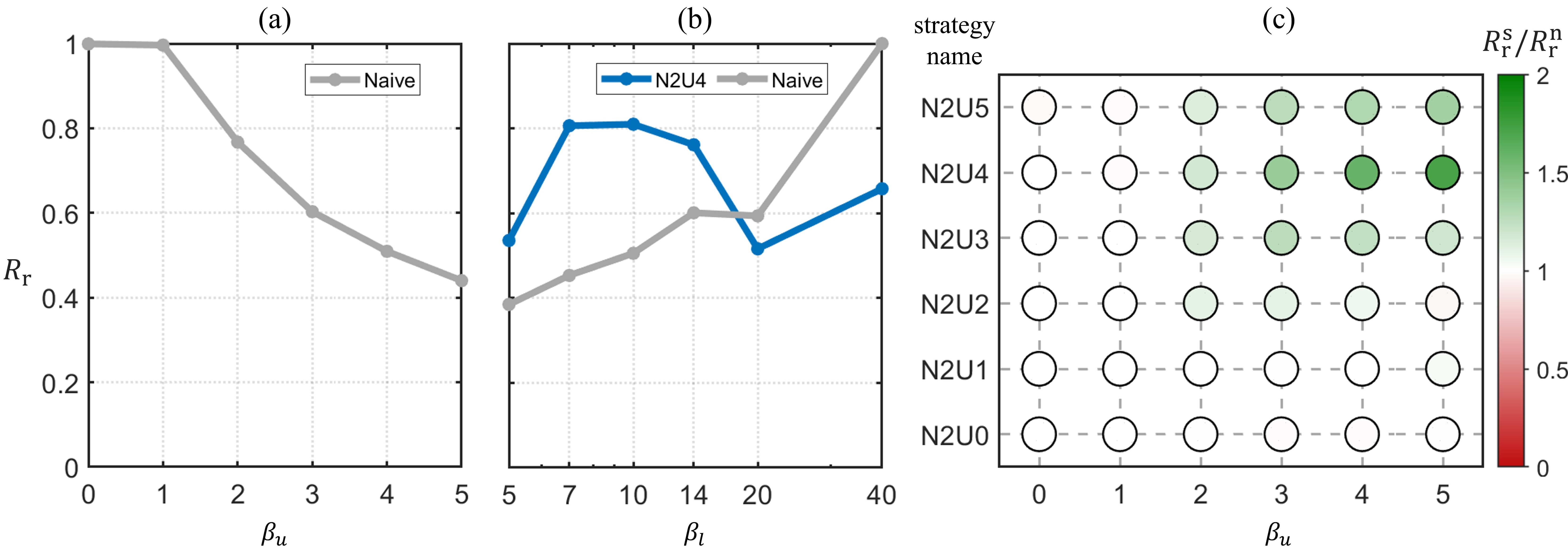}
  \caption{(a) The rendezvous rates of the naive strategy used in different vortex intensities for two-agent swarms while fixing $\beta_l=10$. (b) The rendezvous rates of the naive and the learned N2U4 strategy used in different vortex scales for two-agent swarms while fixing $\beta_u=4$. The horizontal axis is on a logarithmic scale. (c) The performance of six MARL strategies N2U$\beta_u$ trained at vortex intensity $\beta_u$ used in different vortex intensities on the horizontal axis while fixing $\beta_l=10$. The color indicates the $R_\text{r}$ ratio of MARL strategy over the naive strategy used at the same $\beta_u$. The parameter values during testing are shown along the horizontal axis.}
  \label{N2}
\end{figure*}

First, we examine how fluid motion affects the success of rendezvous. The performance of two-agent strategies is evaluated and shown in Fig.~\ref{N2}(a,b,c). Fig.~\ref{N2}(a,b) shows that the rendezvous rate of the naive strategy decreases as $\beta_{u}$ increases or $\beta_{l}$ decreases, indicating greater difficulty in achieving rendezvous. When $\beta_{u} = 0,1$, all swarms rendezvous, because the agents overcome the flow to approach their partners with a propelling speed $v_\text{s}$ large enough compared to the maximal flow speed $u_0$. In addition, Fig.~\ref{sys}(c) shows that with the increase of vortex scale $\beta_{l}$, the flow field becomes less spatially heterogeneous and the velocity gradients are smaller, so that agents at different locations experience more similar fluid velocities $\bm{u}$ and thus the rendezvous task becomes easier. Therefore, the rendezvous problem is highly dependent on the characteristics of the underlying flow field.

The performance of smart rendezvous strategy obtained by MARL training is demonstrated in Fig.~\ref{N2}(c), which shows the rendezvous rate trained and tested under different $\beta_u$ relative to the naive strategy. The diagonal elements in Fig.~\ref{N2}(c) indicates that, when the MARL strategy is trained and tested with the same $\beta_u$, it outperforms the naive baseline, demonstrating the validity of our MARL framework. Notably, N2U4 strategy achieves a 59\% improvement in $R_r$.

We assess the transferability by testing MARL strategy in all considered vortex intensities. As shown in Fig.~\ref{N2}(c), most MARL strategies outperform the naive one in the considered flow intensities. Besides, the performances of both N2U0 and N2U1 strategies are close to the naive strategy even when they are tested in a strong flow of large $\beta_u$, as shown in the lowest two rows of Fig.~\ref{N2}(c). This means that the agents learn a smart strategy similar to the naive strategy when they are trained in a weak-flow environment. In terms of the transferability to different vortex scale, we test an example, i.e., N2U4 strategy, in all considered vortex scales while fixing the intensity as $\beta_u=4$. Fig.~\ref{N2}(b) illustrates that N2U4 strategy is more effective in TGVs of smaller $\beta_l$, indicating the advantage of N2U4 strategy in tackling small-scale structures in flow field. In general, in flow fields with dense vortices, MARL strategy shows strong transferability to flow variations.

\subsection{Performance of multi-agent strategies} 

\begin{figure*}
      \centering
      \includegraphics[width=0.9\textwidth]{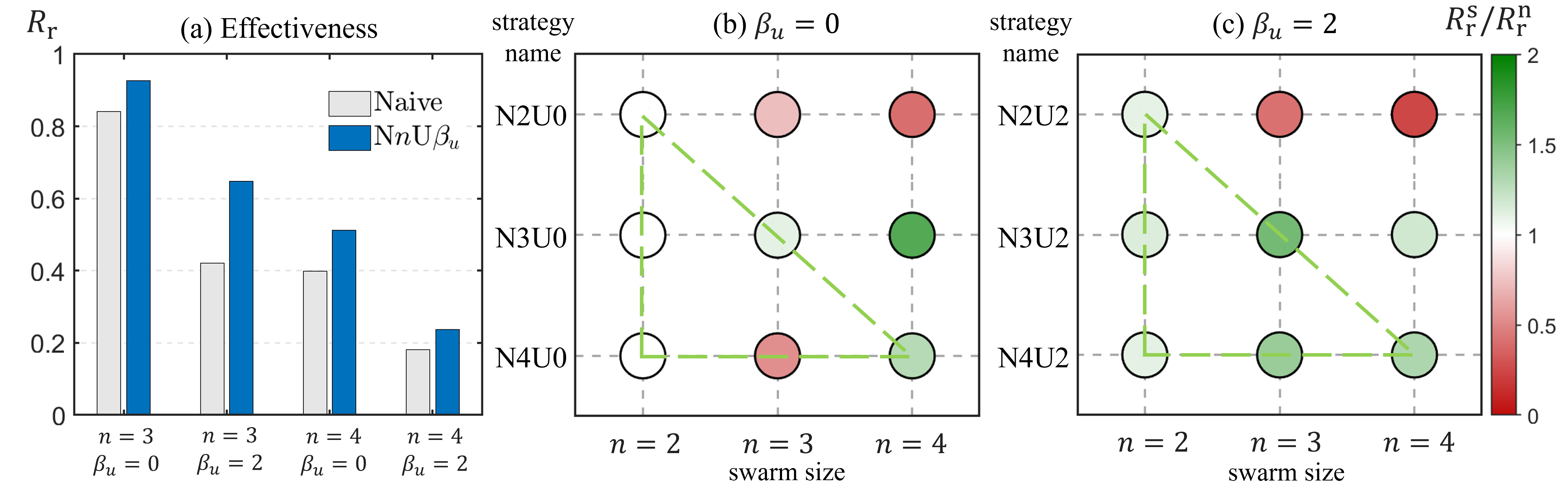}
      \caption{
      (a) The rendezvous rates of strategies for different swarm sizes and in different flow intensities. In the TGVs of (b)$\beta_u=0$ or (c)$\beta_u=2$, the transferability of strategies to the swarm size. The color indicates the $R_\text{r}$ ratio of MARL strategy over the naive strategy used in the case of swarm size $n$.}
      \label{rbn}
\end{figure*}

In this section, we investigate the strategy for swarms of different sizes. The results for swarms of three and four agents in $\beta_u =0,2$ are shown in Fig.~\ref{rbn}(a). The performance of the naive strategy decreases with larger $n$, which means that the rendezvous task for a larger swarm is more difficult. The reason is that the rendezvous requires the distances between any two of the $n$ agents to be smaller than the threshold, which is more difficult for larger swarms because the number of agent pairs is $n(n-1)/2$. Nevertheless, the advantages of MARL strategies over naive strategy demonstrate that our MARL approach is still effective for large swarm.

Our MARL implementation is devised to be independent of swarm size, which means that each strategy can be used in cases where the swarm size is different from that in training. Thus, the strategy transferability to different $n$ is also tested, as shown in Fig.~\ref{rbn}(b,c). The results show that the strategies trained at $n=3$ still outperform the naive strategy when tested for $n=2,4$. The element in the lower triangle in Fig.~\ref{rbn}(b,c), where $n$ during training is larger than that during testing, shows better performance than the upper triangle. The lack of full transferability is not discouraging since transferability is usually not guaranteed in RL problems. To further improve generalization, one could train a single strategy over all considered parameters.

 \subsection{Mechanism of MARL strategy}
\begin{figure*}
  \centering
  \includegraphics[width=1.0\textwidth]{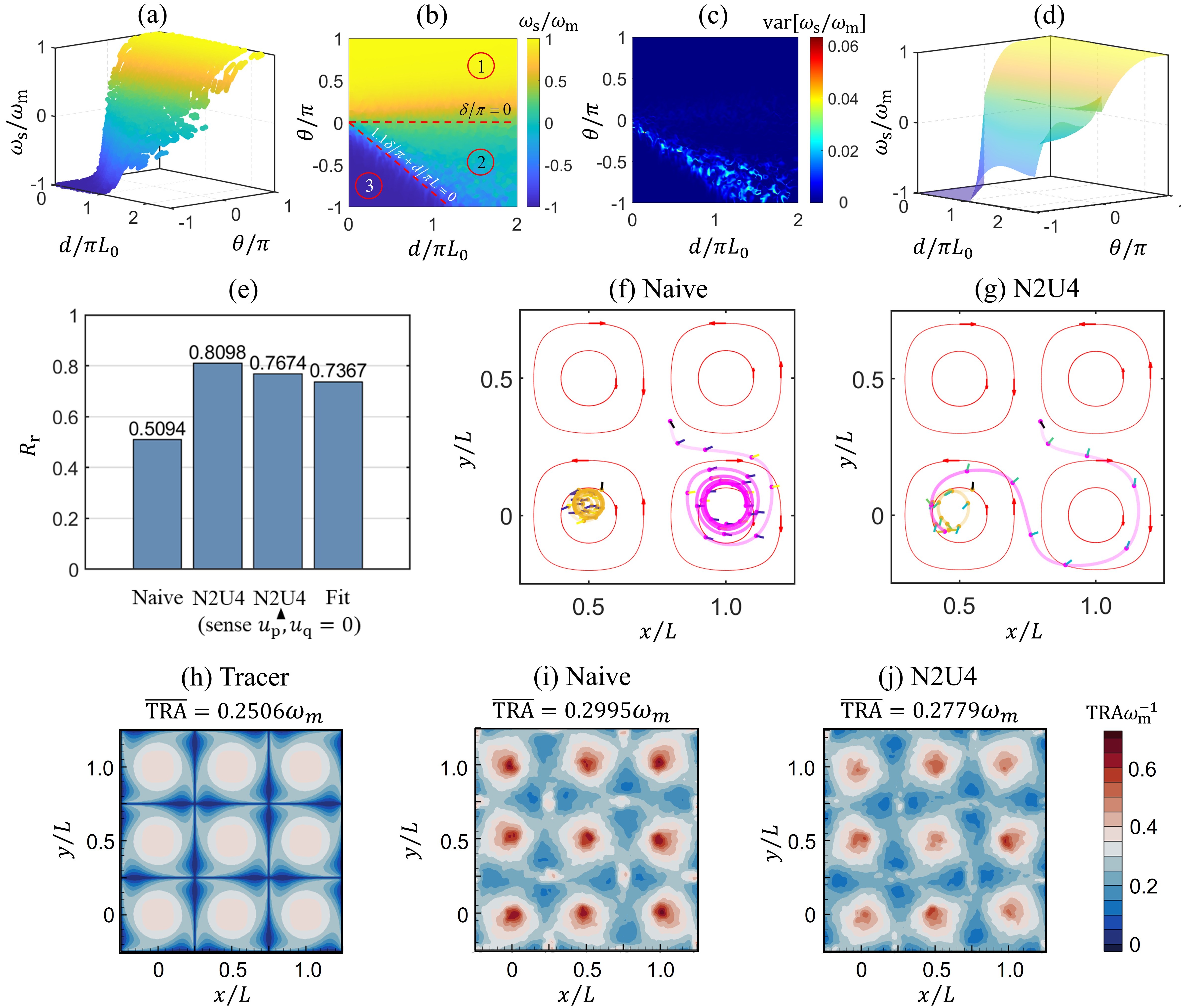}
  \caption{(a) The 3D scatter diagram of state-action samples. (b) The mean action of $\omega_\text{s}$ at different $(d,\theta)$ state. The state regions are divided into three parts by the sharp changes in action (red dashed lines). (c) The variance of action $\omega_\text{s}$ at different $(d,\theta)$ caused by the other states $u_\text{p},u_\text{q}$ instead of network stochasticity, because no network stochasticity exists at action decision. (d) The piecewise function fitted from data in (a). (e) The performance of different state-action maps, where the third bar means the agents take actions by N2U4 strategy but input the state of flow speed $u_\text{p},u_\text{q}$ as zeros. The trajectories of an agent pair moving in TGV of $\beta_u=8$ with (f) naive and (g) N2U4 strategies. Opacity increases with time, and two colors distinguish two agents. Some points on the trajectory are drawn, where the line segments are the tail of the agents, representing orientations. According to the yellow-blue color bar in (b), $\omega_\text{s}$ is described in (a, b, d), and in (f, g), the action $\omega_\text{s}$ is also indicated by the tail color. 
  The TRA distributions for (h) tracer agent whose $v_\text{s}=0$, (i) agent taking naive strategy, (j) agent taking N2U4 strategy moving in TGV of $\beta_u=8$. $\overline{\text{TRA}}$ is the average over the region of $x,y \in[-0.25L,1.25L]$.}
  \label{sa}
\end{figure*}

To understand the mechanism of MARL strategy, we analyze the dependence of action on states by sampling state-action data. Here, we use N2U4 strategy as an example, because N2U4 strategy shows the best performance in Fig.~\ref{N2}(c). For a two-agent swarm, the states include $u_\text{p},u_\text{q},d,\theta$ defined in Fig.~\ref{sys}(e). As seen in Fig.~\ref{sa}(a), the points $(d,\theta,\omega_\text{s})$ approximately form a 2D manifold, suggesting that the control action is primarily governed by $d$ and $\theta$. Therefore, we focus on the dependency of action on $d,\theta$.

Based on the sampled data, we obtain the mean and variance of action over different $u_\text{p}$ and $u_\text{q}$ by Gaussian-weighted gridded statistics (see Appendix \ref{appC}), as shown in Fig.~\ref{sa}(b,c). The mean value of action is the average steering speed at a certain $d,\theta$. In Fig.~\ref{sa}(c), the large variance indicates where action is sensitive to the changes in flow velocities $u_\text{p},u_\text{q}$. Because the regions of large variance occupy only a small portion, the information of the relative position $d, \theta$ dominate the strategy, whereas the local flow velocity $u_\text{p},u_\text{q}$ have a minor effect. This is verified by masking the flow signals $u_\text{p},u_\text{q}$ by zero, which leads to a marginal 4\% decrease in rendezvous rate, as shown in  Fig.~\ref{sa}(e). In other words, N2U4 strategy remains effective even in the absence of local flow information.

To fit the state-action samples with an explicit function, we divide the state region into three parts as shown in Fig.~\ref{sa}(b). Then we obtain a state-action function $S(d/\pi L_0, \theta/\pi)$ through piecewise fitting by third-order polynomials. The function is displayed in Fig.~\ref{sa}(d), and its mathematical expression is in Appendix \ref{appD}. This function provides a heuristic controlling approach for the two-agent rendezvous task, and the corresponding $R_\text{r}$ in Fig.~\ref{sa}(f) proves its effectiveness.

The mechanism of N2U4 strategy is revealed by Fig.~\ref{sa}(a,b). At small $d$, the $\omega_\text{s}$ is an increasing odd function of $\theta$. The same sign of $\omega_{\rm s}$ and $\theta$ indicates that the two agents steer towards each other when they are close, similar to the mechanism of the naive strategy. When $d$ is large, $\omega_\text{s}$ is no longer an odd function of $\theta$. To illustrate the consequence, we show the trajectories of smart agents in Fig.~\ref{sa}(g). When the agent is in the bottom-right vortex, it experiences a state corresponding to region \circled{2} in Fig.~\ref{sa}(b), and it does not steer towards the direction of the other agent. This allows it to escape the bottom-right vortex and later rendezvous with its partner which remains in the lower-left vortex as shown in Fig.~\ref{sa}(g). In contrast, naive agents are trapped in separate vortices as shown in Fig.~\ref{sa}(f).

This vortex-induced trapping effect can be quantitatively verified. We employ Lagrangian coherent structure (LCS) theory, which is a tool for describing transport and mixing processes in fluid systems. This theory has been successfully used to analyze the behaviors of inertial \citep{pengTransportInertialParticles2009}, self-propelled \citep{bermanTransportBarriersSelfpropelled2021,storm2025transport}, and controlled agents \citep{krishnaFiniteTimeLyapunov2023} in flow fields. 
We focus on elliptic Lagrangian Coherent Structure (eLCS) since it detects regions exhibiting sustained coherent rotation. The trajectory rotation average (TRA) is a major indicator in eLCS, which evaluates the rotation intensity of the agent trajectory \citep{hallerQuasiobjectiveCoherentStructure2021}. TRA is expressed as:
\begin{equation}
\begin{aligned}
\text{TRA}(\bm{x}(t_0)) = \frac{1}{t_N-t_0}\sum_{i=0}^{N-1}\cos^{-1}\frac{<\bm{v}(t_i),\bm{v}(t_{i+1})>}{|\bm{v}(t_i)||\bm{v}(t_{i+1})|},
\end{aligned}
\end{equation}
where $\bm{x}(t),\bm{v}(t)$ represent the position and velocity vector of the agent at time $t$. The time window $t_N-t_0$ is divided into $N$ terms, and in each term, the angle variation between the initial and next velocity vectors $\bm{v}(t_i),\bm{v}(t_{i+1})$ is calculated. The sum of angle variations over the time window represents the rotation of the trajectory started at $\bm{x}(t_0)$. Here, we set $N=20$ and $t_N-t_0=5\tau$. Fig.~\ref{sa}(h) shows large TRA of ambient flow motion, obtained from trajectories of tracer agents which are consistent with streamlines. Moreover, comparing Fig.~\ref{sa}(i, j), we can see that TRA for the naive agent is larger than that for N2U4 agent, especially in vortices with negative vorticity, as indicated in Fig.~\ref{sa}(g). This coincides with the mechanism of avoiding trapping in a negative vortex analyzed in the previous paragraph.

From the above analysis, we can conclude that the non-trapping mechanism of MARL strategy is powerful in environments with dense vortices. This mechanism explains why MARL strategy shows less advantage over naive strategy with the increase of $\beta_l$ as shown in Fig.~\ref{N2}(d).

\subsection{Rendezvous distributions}\label{sec:dist}
To further examine these findings, we analyze the rendezvous behaviors in a statistical way to understand the flow effects. We define $P(x,y)$ as the probability density of a swarm, randomly initialized in region $\Phi$, that successfully reach rendezvous in one episode at position $(x,y)$. For the two-agent naive and MARL strategies, $P(x,y)$ is obtained by Gaussian kernel density estimation (see Appendix \ref{appA}), as shown in Fig.~\ref{dtb}. The distributions of $P$ of naive strategy in different flow intensities are shown in Fig.~\ref{dtb}(a, b). The largest $P$ is found at the center of $\Phi$, where the two agents, initialized at random positions, move statistically equal distances to meet. When flow is present (Fig.~\ref{dtb}(b)), agents rendezvous near vortex centers, indicated by the large $P$ in vortical regions. Comparing Fig.~\ref{dtb}(b) with (d), it is found that N2U4 strategies increase $P$, which explains the higher $R_\text{r}$ achieved by MARL strategies displayed in Fig.~\ref{N2}(a).

\begin{figure*}
  \centering
  \includegraphics[width=0.96\textwidth]{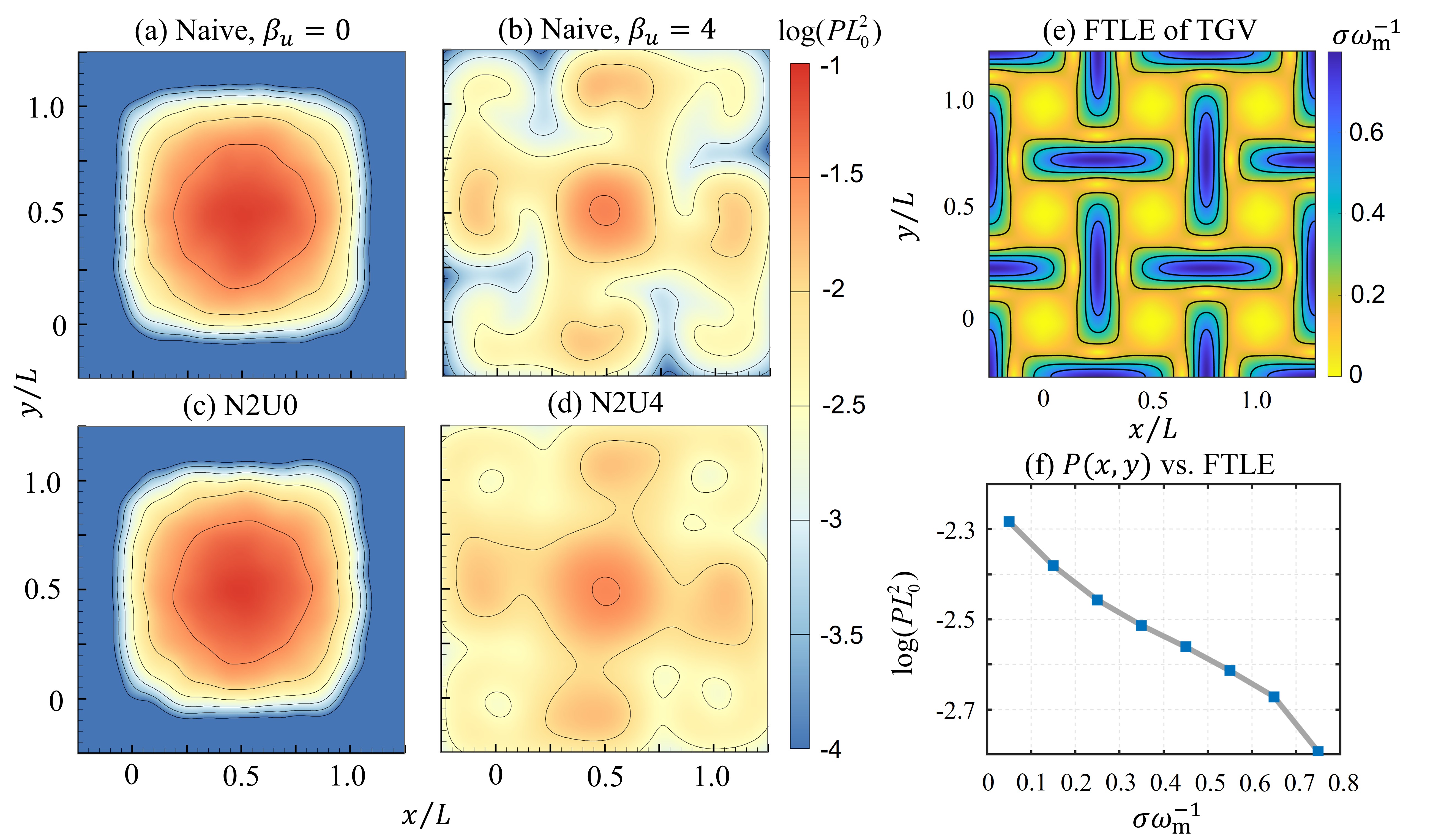}
  \caption{The distributions of rendezvous probability density $P(x,y)$ for (a-b) naive strategy and (c-d) MARL strategies used in the training TGVs, where color indicates the logarithm of $P(x,y)$. (e) The FTLE $\sigma$ distribution of TGV of $\beta_u=4$. (f) The statistical correlation between rendezvous probability $P(x,y)$ and FTLE $\sigma$.}
  \label{dtb}
\end{figure*}

To explain the correlation between rendezvous distribution and the flow structure, we analyze the distance evolution of two agents caused by fluid motion by considering the finite-time Lyapunov exponent (FTLE), which is widely used to describe the hyperbolic LCS \cite{hallerLagrangianCoherentStructures2014}. With the flow map $\bm{F}^t_{t_0}(\bm{x_0})= \bm{x}(t;t_0,\bm{x_0})$ representing position evolution of a tracer agent starting at time $t_0$ and position $\bm{x_0}$, the evolution of the relative position of a pair of agents $\bm{r}(t)=\nabla \bm{F}^t_{t_0}(\bm{x_0})\bm{r_0}$. The magnitude of $\bm{r}(t)$ is:
\begin{equation}
\begin{aligned}
|\bm{r}(t)|^2&= [\nabla \bm{F}^t_{t_0}(\bm{x_0})\bm{r_0}]^T\nabla \bm{F}^t_{t_0}(\bm{x_0})\bm{r_0}\\&=\bm{r_0}^T[\nabla \bm{F}^t_{t_0}(\bm{x_0})]^T\bm{F}^t_{t_0}(\bm{x_0})\bm{r_0}\\&=\bm{r_0}^T\bm{C}\bm{r_0},
\end{aligned}
\end{equation}
where $\bm{C}=[\nabla \bm{F}^t_{t_0}(\bm{x_0})]^T\nabla\bm{F}^t_{t_0}(\bm{x_0})$ is the Cauchy-Green tensor. With the eigenvalues $\lambda_i$ of $\bm{C}$, the FTLE value is:
\begin{equation}
\sigma=\frac{\log{\sqrt{\max(\lambda_i)}}}{t-t_0}.
\end{equation}

Since $\bm{C}$ is a symmetric matrix, the Rayleigh quotient $R=\frac{\bm{r_0}^T\bm{C}\bm{r_0}}{\bm{r_0}^T\bm{r_0}}=\frac{|\bm{r}(t)|^2}{\bm{r_0}^T\bm{r_0}}$ describes the stretching of agent distance, which satisfies $R_{\max} = \max\lambda_i$. Hence, $|\bm{r}_{\max}(t)|= |\bm{r}(t_0)|e^{\sigma(t-t_0)}$. In the two-dimensional system, the fluid is incompressible, thus $\lambda_{\max}\lambda_{\min}=\det{\bm{C}}=1$ so that $|\bm{r}_{\min}(t)|=|\bm{r}(t_0)|e^{-\sigma(t-t_0)}$. Physically, a large FTLE measures strong deformation of a fluid particle, which leads to exponentially increasing separation but limited compression of adjacent agents in the fluid particle. Therefore, it is difficult for agents to rendezvous if they experience spaces with large FTLE.

We calculate FTLE distribution for TGV of $\beta_u=4, \beta_l=10$, with the end time $t_E-t_0 = -2.5\tau$. By integrating trajectories backward in time, we can find whether agents experienced strong deformation before finally rendezvousing at position $\bm{x_0}$. The FTLE distribution is displayed in Fig.~\ref{dtb}(e), which presents a negative correlation with rendezvous probability in Fig.~\ref{dtb}(b). Mean $\log(P(x,y)L_0^2)$ of different FTLE is shown in Fig.~\ref{dtb}(b). The presence of negative correlation verifies the above theoretical analysis.

Although FTLE is calculated from a Lagrangian property of the flow, it still reflects Eulerian characteristics of the flow. Comparing Fig.~\ref{sys}(a) and Fig.~\ref{dtb}(b), we can see that high FTLE occurs at the vortex edge, where Eulerian deformation is strong, corresponding to a small $Q$ value.

The connection between FTLE and the Eulerian deformation can be understood as follows.
From an Eulerian perspective, the evolution of the separation between two agents satisfies $\dot{\bm{r}}(t)=\nabla \bm{u}(\bm{x},t)\bm{r}(t)$. Meanwhile, the deformation gradient obeys $\nabla \bm{\dot{F}}^t_{t_0}(\bm{x_0}) = \nabla \bm{u}(\bm{x},t)\nabla \bm{F}^t_{t_0}(\bm{x_0})$, which yields $\dot{\bm{C}} =[\nabla\bm{F}^t_{t_0}(\bm{x_0})]^T \bm{E}\nabla \bm{F}^t_{t_0}(\bm{x_0})$, where Eulerian deformation tensor of fluid $\bm{E} = [\nabla \bm{u}(\bm{x},t)]^T +\nabla \bm{u}(\bm{x},t)$.
The Cauchy-Green tensor $\bm{C}$ is initially the identity tensor, corresponding to $\sigma=0$. As the deformation induced by $\bm{E}$ accumulates along trajectories, the eigenvalues of $\bm{C}$ separate, leading to an increase in the FTLE.

Therefore, large FTLE values are associated with trajectories that experience strong accumulated Eulerian deformation. Since large FTLE is correlated with a small rendezvous probability (Fig.~\ref{dtb}(f)), lower rendezvous rates are expected near vortex edges, where the Eulerian deformation is strong. Moreover, the rendezvous task becomes increasingly difficult for larger $\beta_u$ or smaller $\beta_l$, as discussed in Sec.~\ref{sec:2agent}, because the magnitude of $\bm{E}$ scales with $\beta_u/\beta_l$.

\section{Conclusions}
In this study, we have demonstrated that a multi-agent reinforcement learning approach is highly effective in solving the rendezvous problem under the influence of ambient fluid motions. The strategies yielded through the DS-PPO framework significantly outperform naive baselines across various flow intensities and swarm sizes. These results highlight the capability of MARL in exploring swarm intelligence within complex flow environments where traditional control policies often remain suboptimal.

The superiority of MARL strategy, specifically N2U4, is attributed to its capacity to mitigate 'vortex trapping', which inherently limits naive strategies. While naive agents tend to get trapped in separate vortices, the trained RL agents can escape trapping by a symmetry-breaking mechanism in the state-action map. This behavior is quantitatively supported by the TRA analysis, which reveals that RL agents exhibit weaker rotation than naive ones. This suggests that MARL does not merely optimize paths but evolves environmental intelligence to form beneficial coherent structures.

Furthermore, by analyzing the statistical distributions of successful rendezvous, we uncover how local fluid kinematics govern agent coordination. We demonstrate that fluid deformation is an inherent obstacle: regions with high finite-time Lyapunov exponents (FTLE) exponentially drive agents apart. These findings provide a crucial practical guideline, strongly suggesting that future multi-agent rendezvous targets should be strategically planned in regions exhibiting weak fluid deformation.

Based on these understanding, more detailed fluid effects can be further interpreted by integrating high-fidelity methods such as the immersed boundary method \citep{peskinImmersedBoundaryMethod2002}, to account for precise agent-fluid and agent-agent interactions. Additionally, exploring the effects of communication limitations, sensory noise, and physical obstacles will be crucial for the application of these strategies to oceanic or atmospheric vehicles. Beyond rendezvous, the framework demonstrated here holds significant potential for other homogeneous collective tasks, such as flocking or collaborative search in turbulent environments.

\begin{acknowledgments}
B. L. and L. Z. acknowledge the support from National Science Foundation of China (Grants No. 12472224, 27892252104 and 12388101). J. Q. acknowledges the support from Vetenskapsrådet (Grants No. 2018-03974 and 2023-03617).
\end{acknowledgments}

\appendix
\section{Gaussian kernel density estimation on rendezvous distribution}\label{appA}
We define the rendezvous probability density $P(x,y)$ that satisfies:
\begin{equation}
 \begin{gathered}
   R_\text{r}(A)  = \iint_{A} P(x,y)\mathrm{d}x\mathrm{d}y,
  \end{gathered}
  \label{eq:pdf}
\end{equation}
where $R_\text{r}(A)$ denotes the probability that a swarm, randomly initialized in $\Phi(x,y\in[0,L])$, rendezvous at the position inside the area $A$. With $(x_i,y_i)$ describing the center position of swarm $i$ when the swarm is judged to rendezvous, the rendezvous probability densities $P(x,y)$ can be obtained with Gaussian kernel density estimation method as:
\begin{equation}
 \begin{gathered}
    P(x,y) = \frac{1}{N h_x h_y} \sum_{i=1}^{N_\text{r}} K\left(\frac{x-x_i}{h_x}\right) K\left(\frac{y-y_i}{h_y}\right),\\
    K(m)=\frac{1}{\sqrt{2\pi}}\exp(-m^2/2),\\
  \end{gathered}
  \label{gkde}
\end{equation}
where $K(m)$ is the Gaussian kernel function, and $h_x,h_y$ are the bandwidths selected with Silverman \cite{silvermanDensityEstimationStatistics2018} rule.

\section{Gaussian-weighted gridded statistics}\label{appC}
We again use Gaussian kernel function in \eqref{gkde} to obtain action statistics in the state space. This approach allows for continuous field reconstruction from numerous discrete data points. Given a set of data samples $(x_{i}, y_{i}, z_i)$, we can quantify the mean and variance of action $\bar{z}$, $\text{var}[z]$ at point $(x, y)$ as follows: 
\begin{equation}
\begin{gathered}     
\bar{z}(x,y) = \frac{\sum_{\mathcal{R}} K(\frac{x-x_{i}}{h})K(\frac{y-y_{i}}{h}) z_i}{\sum_\mathcal{R}K(\frac{x-x_{i}}{h})K(\frac{y-y_{i}}{h})},\\     
\text{var}[z](x,y) = \frac{\sum_{\mathcal{R}} K(\frac{x-x_{i}}{h})K(\frac{y-y_{i}}{h}) (\bar{z} - z_i)^2}{\sum_\mathcal{R} K(\frac{x-x_{i}}{h})K(\frac{y-y_{i}}{h})},\\
\end{gathered}
\end{equation}
where $\mathcal{R}$ represents the region of $|x_{i}-x|<H,|y_{i} - y|<H$. Here, we set $H=40h, h=0.01$. In our work, $x,y,z$ correspond to the states $d/\pi L_0,\theta/\pi$ and action $\omega/\omega_\text{m}$, respectively, to analyze the dependencies of the action on these two states, as shown in Fig.~\ref{sa}(b,c).

\section{Fitting the map from state to action}\label{appD}
The explicit state-action function $S(\hat{d},\hat{\theta})$ is fitted by third-order polynomials with the least-square method, where $\hat{d}=d/\pi L_0,\hat{\theta}=\theta/\pi$. In addition, the function is bounded to [-1,1], and we use the clip function to restrict it. The expression of $S(\hat{d},\hat{\theta})$ is:

\begin{equation}
S(\hat{d},\hat{\theta}) = 
\begin{cases}
\text{clip}[f_1(\hat{d},\hat{\theta})], & \text{if } \hat{\theta} > 0 \\
\text{clip}[f_2(\hat{d},\hat{\theta})], & \text{if } \hat{\theta} \leq 0\text{ and }  \hat{d} + 1.1\hat{\theta} \geq 0 \\
-1, & \text{if } \hat{d} + 1.1\hat{\theta} < 0
\end{cases}
\end{equation}

\begin{equation}
\begin{aligned}
f_1(\hat{d},\hat{\theta}) &=-0.16\hat{d}^3\hat{\theta}^3 + 0.22\hat{d}^3\hat{\theta}^2 - 0.10\hat{d}^3\hat{\theta} \\
&+ 0.02\hat{d}^3+1.27\hat{d}^2\hat{\theta}^3- 1.77\hat{d}^2\hat{\theta}^2 + 0.70\hat{d}^2\hat{\theta}\\
&- 0.07\hat{d}^2 -3.20\hat{d}\hat{\theta}^3 + 4.64\hat{d}\hat{\theta}^2- 1.75\hat{d}\hat{\theta}\\
&+ 0.07\hat{d}+3.07\hat{\theta}^3 - 5.76\hat{\theta}^2 + 3.55\hat{\theta} + 0.26,\\
\end{aligned}
\end{equation}

\begin{equation}
\begin{aligned}
f_2(\hat{d},\hat{\theta}) &= 2.21\hat{d}^3\hat{\theta}^3 + 2.25\hat{d}^3\hat{\theta}^2 - 0.51\hat{d}^3\hat{\theta} + 0.14\hat{d}^3\\
& -8.84\hat{d}^2\hat{\theta}^3- 9.71\hat{d}^2\hat{\theta}^2 + 3.06\hat{d}^2\hat{\theta} - 0.50\hat{d}^2 \\
&+13.00\hat{d}\hat{\theta}^3 + 17.79\hat{d}\hat{\theta}^2- 3.95\hat{d}\hat{\theta}+ 0.55\hat{d}\\
& -5.92\hat{\theta}^3 - 10.88\hat{\theta}^2 + 1.96\hat{\theta} + 0.11.\\
\end{aligned}
\end{equation}

The fitted function is shown in Fig.~\ref{sa}(d).

\bibliography{apssamp}

\end{document}